\documentclass[a4paper,10pt]{article}

\usepackage{amsmath}
\usepackage{amsthm}
\usepackage{amsfonts}
\usepackage[linesnumbered,ruled]{algorithm2e}

\newtheorem{lemma}{Lemma}[section]
\newtheorem{corollary}[lemma]{Corollary}

\newtheorem{theorem}[lemma]{Theorem}

\newtheorem{example}{Example}[section]

\begin{document}

\title{Minimizing Symmetric Set Functions Faster}
\author{Michael Brinkmeier\\[3mm]{\small Institute for Theoretical Computer Science}\\
{\small Technical University Ilmenau, Ilmenau, Germany}\\
{\small\tt mbrinkme@tu-ilmenau.de}}



\maketitle

\begin{abstract}
We describe a combinatorial algorithm which, given a monotone and consistent symmetric set function
$d$ on a finite set $V$ in the sense of Rizzi \cite{rizzi00symmetric}, constructs a non trivial set
$S$ minimizing $d(S,V\setminus S)$. This includes the possibility for the minimization of symmetric
submodular functions. The presented algorithm requires at most as much time as the one in
\cite{rizzi00symmetric}, but depending on the function $d$, it may allow several improvements.

\bigskip

\parindent0pt
{\em Keywords:} symmetric set function, symmetric submodular function, minimum cut, optimal bipartition
\end{abstract}

\section{Introduction}

Let $V$ be a finite set. A {\em symmetric set function $d$} is a map assigning a real number
to each pair $(S,T)$ of {\em disjoint} subsets of $V$ satisfying $d(S,T) = d(T,S)$. The function
$d$ is called {\em monotone}, if for every pair $S,T$ of disjoint sets and $T'\subseteq T$
the inequality $d(S,T') \leq d(S,T)$ holds. It is called {\em consistent}, if for every
triple $R,S,T$ of pairwise disjoint set $d(S,R) \geq d(T,R)$ implies $d(S,R\cup T) \geq
d(S\cup R,T)$. As the examples below indicate, the construction of a {\em minimum bipartition}
of $(V,d)$, ie. a nontrivial\footnote{{\em Nontrivial} means $\emptyset\neq S\neq V$} subset $S$ of $V$
minimizing $d(S,V\setminus S)$, is important for many applications.

\begin{example}\label{exp:mincut}
If $G=(V,E)$ is a (hyper-)graph with weighted edges, the weight $w(S,T)$ of edges between
two disjoint sets of vertices is a monotone and consistent symmetric set function.
A minimum bipartition of $(V,d)$ is a minimum cut of the (hyper-)graph $G$.
\cite{nagamochi92connectivity,nagamochi94implementing,frank94edge,stoer97mincut,klimmek95simple,brinkmeier05mincut}
\end{example}

\begin{example}\label{exp:submod}
If $f\colon 2^V\to \mathbb{R}$ is {\em submodular}, ie.
$$ f(S) + f(T) \geq f(S\cup T) + f(S\cup T) $$
for all subsets $S,T\subseteq V$, then the {\em generalized connectivity 
function}
$$ c_f(S,T) := f(S) + f(T) -f(S\cup T)$$
is a monotone and consistent symmetric set function.
If $f$ is symmetric, ie. $f(A)=f(V\setminus A)$, then a minimum bipartition of 
$c_f$ is a nontrivial set minimizing $f$.
\cite{cunningham92decomposition,queyranne98minimizing,rizzi00symmetric}
\end{example}

A local version of the minimum bipartition problem, consists in the detection of a set $S$, such that
$d(S,V\setminus S)$ is minimal among all sets {\em separating} two given elements $s$ and $t$,
ie. $s\in S$ and $t\not\in S$. We define
$$\lambda_{(V,d)}(s,t) = \min\left\{ d(S,V\setminus S) \mid s\in S, t\not\in S\right\} $$
and
$$\lambda_{(V,d)} = \min\left\{ \lambda_{(V,d)}(s,t) \mid s,t\in V \right\}. $$
Hence, we want to find a set $S$, such that $d(S,V\setminus S) = \lambda_{(V,d)}$.

Based on the work of Nagamochi and Ibaraki \cite{nagamochi92lineartime,nagamochi92connectivity},
Stoer and Wagner in \cite{stoer97mincut}, and independently Frank in \cite{frank94edge}, described
an algorithm constructing a minimum cut of a weighted graph (exp. \ref{exp:mincut}). This algorithm
was generalized by Queyranne
\cite{queyranne98minimizing} to the minimization of {\em symmetric submodular functions}
(exp. \ref{exp:submod}), and subsequently by Nagamochi and Ibaraki \cite{nagamochi98submodular}
to a wider class of functions, satisfying a less restrictive symmetry condition.

Another generalization was given by Rizzi in \cite{rizzi00symmetric}. He proved that the algorithm
of Stoer/Wagner/Frank/Queyranne generalizes to the minimization of monotone and consistent symmetric
set functions. Based on his work, we will describe a variation of the algorithm which provides
possibilities to reduce the required time. Unfortunately, these improvements depend heavily on the function
to be minimized. But we can guarantee, that the new algorithm requires at most the same time as that
one described in \cite{rizzi00symmetric}.

One example for the possible improvement can be found in \cite{brinkmeier05mincut}. There, the author
used the techniques presented in this paper to reduce the running time of the algorithm of Stoer and
Wagner for the minimum cut of an integer weighted graph from $O(nm+n^2\log n)$ to $O(\delta n^2)$,
with $\delta$ being the minimum degree\footnote{`Reduction' is a bit dangerous, since the running time 
is $O(\delta n^2)$ and hence pseudo polynomial. Nonetheless, if the edge weights are reasonable, one can
expect the new algorithm to be faster than the old one.}.

We proceed as follows. First, we introduce {\em lax-back orders}, generalizing the max-back
orders required by Rizzi in \cite{rizzi00symmetric}. Following that, we prove that the lax-back
orders provide the possibility to identify several elements, hence reducing the size of $V$,
without losing optimal sets. Then these results are used to describe and prove a new algorithm,
bearing several possible improvements of the ones given in \cite{rizzi00symmetric} and 
\cite{queyranne98minimizing}.

\section{Lax-Back Orders}

In the following, for each finite set $V$ we write $n=|V|$ and a singleton set $\{v\}\subseteq V$ will
simply be denoted by $v$.

Before we proceed, we make a simple and useful observation.
\begin{lemma}\label{lem:triangle}
Let $d$ be a symmetric set function on $V$ and $u,v,w$ elements in $V$ and $\tau\in\mathbb{R}$, then
$\lambda_{(V,d)}(u,v), \lambda_{(V,d)}(v,w) \geq \tau$ implies $\lambda_{(V,d)}(u,w) \geq\tau$.
\end{lemma}
\begin{proof}
Let $S$ be a subset of $V$ with $u\in S$ and $w\not\in S$, such that $\lambda_{(V,d)}(u,w) = d(S,V\setminus S)$.
If $v\in S$, $S$ separates $v$ and $w$ and hence $d(S,V\setminus S) \geq \lambda_{(V,d)}(v,w)\geq\tau$.
If $v\not\in S$ we have $d(S,V\setminus S) \geq \lambda_{(V,d)}(u,v)\geq\tau$.
\end{proof}

Let $V$ be a finite set and $d$ a symmetric set function on $V$. An ordered pair $(s,t)$ of elements of
$V$ is called {\em good}, if  $d(t,V\setminus t)=\lambda_{(V,d)}(s,t)$.
As Rizzi proved in \cite{rizzi00symmetric}, such a pair $(v_{n-1},v_n)$ can be found by constructing a
total order $v_1,\dots, v_n$ on $V$, such that 
$$ d(v_i,\{v_1,\dots, v_{i-1}\}) \geq d(v_j, \{v_1,\dots,v_{i-1}\}) \quad \text{ for } 
1 \leq i < j\leq n.$$
An order of this type is called {\em max-back order} for $(V,d)$.

\begin{lemma}[Rizzi \cite{rizzi00symmetric}]\label{lem:rizzi}
Let $v_1,\dots, v_n$ be a max-back order for $(V,d)$. Then $(v_{n-1},v_n)$ is
good for $(V,d)$.
\end{lemma}

Instead of using the original function $d$, we adapt it and introduce a {\em threshold $\tau$}.
\begin{lemma}\label{lem:lax.symm}
Let $d$ be a monotone and consistent symmetric set function and $\tau$ a real number.
Then $\hat d(S,T) = \min\left\{ \tau, d(S,T)\right\}$ is a monotone and consistent symmetric
set function.
\end{lemma}
\begin{proof}
Obviously $\hat d$ is symmetric. Now let $S,T$ be two disjoint sets and $T'\subseteq T$. Then we have
$$ \hat d(S,T') = \min\{\tau, d(S,T')\} \leq \min\{\tau, d(S,T)\} = \hat d(S,T). $$
Hence, $\hat d$ is monotone.

Now let $R,S,T$ be three pairwise disjoint sets such that $\hat d(S,R) \geq \hat d(T,R)$.
If $d(S,R) < \tau$, then $d(T,R) \leq d(S,R)$. Due to the consistency of $d$,
this leads to
$$ \hat d(S,R \cup T) = \min\left\{ \tau, d(S,R \cup T) \right\}
\geq \min\left\{ \tau, d(S\cup R,T)\right\} = \hat d(S\cup R,T). $$
If on the other hand $d(S,R) \geq \tau$, we have
$$\hat d(S,R\cup T) = \min\{\tau, d(S,R\cup T)\} \geq \min\{\tau, d(S,R)\} = \tau.$$
Furthermore,
\begin{align*}
\hat d(S\cup R,T) 
&= \min\left\{\tau, d(S\cup R,T) \right\} \\
&= 
\begin{cases}
\tau & \text{ if } d(S\cup R,T) \geq \tau \\
d(S\cup R,T) & \text{ if } d(S\cup R,T) < \tau \\
\end{cases}\\
&\leq \tau \leq \hat d(S, R\cup T),
\end{align*}
concluding the proof of the consistency of $\hat d$.
\end{proof}

A max-back order $v_1,\dots ,v_n$ for $(V,\hat d)$ satisfies
$$ \min\left\{ \tau, d(v_i,\{v_1,\dots, v_{i-1}\})\right\} \geq \min\left\{\tau, 
d(v_j, \{v_1,\dots,v_{i-1}\})\right\} \quad \text{ for } 
1 \leq i < j\leq n.$$
Such an order is called {\em lax-back order with threshold $\tau$} for $(V,d)$.

\begin{corollary}\label{cor:lax}
Let $d$ be a symmetric set function on $V$ and $v_1,\dots, v_n$ a lax-back order with threshold
$\tau$ for $(V,d)$. Then the pair $(v_{n-1},v_n)$ is {\em $\tau$-good} for $(V,d)$, ie.
for each set $T$ with $v_n\in T$ and $v_{n-1}\not\in T$, we have
$$ \min\left\{\tau, d(\{v_n\},V\setminus\{v_n\}) \right\} = \min\left\{ 
\tau, \lambda_{(V,d)}(v_n,v_{n-1})\right\}.$$
\end{corollary}

The proof of the following lemma is straightforward.
\begin{lemma}\label{lem:monotony}
Let $d$ be a monotone and consistent symmetric set function on $V$.
\begin{enumerate}
\item If $v_1,\dots, v_n$ is a max-back order for $(V,d)$, then it is a lax-back order
with threshold $\tau$ for every $\tau$.
\item If $v_1, \dots, v_n$ is a lax-back order for $(V,d)$ with threshold $\tau$, then
it is a lax-back order with threshold $\tau'$ for any $\tau' \leq \tau$.
\end{enumerate}
\end{lemma}

As the preceeding lemma shows, we can interpret a max-back order as a lax back-order
with threshold $\infty$.

\section{Contraction and Lax-Back Orders}

%

In the following let $d$ be a fixed monotone and consistent symmetric set function on $V$.
In \cite{rizzi00symmetric} Rizzi used a max-back order for $(V,d)$ to identify one good pair
$(v_{n-1},v_n)$. He then {\em identified} these two elements, obtaining a monotone and consistent
symmetric set function $d'$ on a smaller set $V'$. This process of {\em identification} or
{\em contraction} may easily be extended to arbitrary partitions of $V$.

\newcommand{\cR}{\mathcal{R}}
\newcommand{\cS}{\mathcal{S}}
\newcommand{\cT}{\mathcal{T}}
\newcommand{\cV}{\mathcal{V}}

Let $\cV = \left\{ U_1,\dots, U_k\right\}$ be an arbitrary partition of $V$. For each element $v\in V$, the
{\em class} of $v$ is denoted by $[v]$, ie. $[v]=U_i$ for $v\in U_i$. 
If, on the other hand, $\cS$ is a set of
classes of $\cV$, we write $$\cup\cS = \bigcup\limits_{U_i\in \cV} U_i \subseteq V$$
for the set of members of classes in $\cS$.

If $d$ is a symmetric set function, we define the {\em induced function $d_\cV$} on $\cV$ as
$$ d_\cV(\cS,\cT) := d\left( \cup\cS,\cup\cT\right). $$
It is easy to check, that $d_\cV$ is monotone and consistent if $d$ is.

%

\begin{lemma}\label{lem:sub}
Let $d$ be a monotone and consistent symmetric set function on $d$. If $v_1,\dots, v_n$ is a lax-back-order 
with threshold $\tau$ for $(V,d)$, then 
$$\min\left\{\tau, \lambda_{(V,d)}(v_{i-1},v_i)\right\} \geq \min\left\{\tau, 
d(v_i,\{v_1,\dots, v_{i-1}\})\right\}$$
for $2\leq i \leq n$.
\end{lemma}
\begin{proof}
Obviously, the restriction $d_i$ of $d$ to $V_i := \{v_1,\dots v_i\}$ is a monotone and consistent symmetric
set function on $V_i$ and hence $v_1,\dots, v_i$ is a lax-back-order with threshold $\tau$ for $(V_i,d_i)$.
By corollary \ref{cor:lax} we have
$$ \min\left\{\tau, d_i(v_i,V_{i-1})\right\} = \min\left\{\tau,\lambda_{(V_i,d_i)}(v_{i-1},v_i)\right\}. $$
Now let $S$ be an arbitrary subset of $V$ with $v_i\in S$ and $v_{i-1}\not\in S$. Then $S_i := S\cap V_i$
separates $v_i$ and $v_{i-1}$ in $V_i$ and hence, due to the monotony of $d$,
$$ \lambda_{(V_i,d_i)}(v_i,v_{i-1}) \leq d_i(S_i,V_i\setminus S_i) = d(S_i,V_i\setminus S_i)
\leq d(S,V\setminus S), $$
and thus
$$ \lambda_{(V_i,d_i)}(v_i,v_{i-1}) \leq \lambda_{(V,d)}(v_i,v_{i-1}).$$
In combination with the observation made above, this leads to
$$ \min\left\{\tau, d_i(v_i,V_{i-1})\right\} = \min\left\{\tau,\lambda_{(V_i,d_i)}(v_{i-1},v_i)\right\} 
	\leq \left\{\tau, \lambda_{(V,d)}(v_i,v_{i-1})\right\} .$$
\end{proof}

Let $v_1,\dots, v_n$ be a lax-back order with threshold $\tau$ for $(V,d)$.
If $\lambda_{(V,d)} < \tau$, then no pair $(v_{i-1},v_i)$ of elements with
$d(v_i,V_{i-1}) \geq \tau$ can be separated by a set $S$ with $d(S,V\setminus S) = \lambda_{(V,d)}$.
Hence we may identify $v_i$ and $v_{i-1}$, without increasing $\lambda_{(V,d)}$. More precisely,
we have the following result.

\begin{lemma}\label{lem:ident}
Let $d$ be a monotone and consistent symmetric set function on $V$ and $\tau$.
If $\cV$ is a partition of $V$, satisfying  
$\lambda_{(V,d)}(u,v) \geq \tau$ for each pair $u,v$ of elements with $[u]=[v]$,
then for each pair $s,t$ of elements of $V$ with $[s]\neq [t]$, we have
$$ \min\left\{\tau, \lambda_{(\cV,d_\cV)}(s,t)\right\} = \min\left\{ \tau, \lambda_{(V,d)}(s,t)\right\},$$
and
$$ \min\left\{\tau, \lambda_{(\cV,d_\cV)}\right\} = \min\left\{ \tau, \lambda_{(V,d)}\right\}.$$
\end{lemma}
\begin{proof}
Let $\cS$ be a subset of $\cV$ with $[s]\in \cS$ and $[t]\not\in\cS$. For $S=\cup\cS$ we have $d_\cV(\cS,\cV\setminus\cS) =d(S,V\setminus S)$, and thus 
$\lambda_{(\cV,d_\cV)}([s],[t]) \geq \lambda_{(v,d)}(s,t)$, resulting in
$$ \min\left\{\tau, \lambda_{(\cV,d_\cV)}\right\} \geq \min\left\{ \tau, \lambda_{(V,d)}\right\}.$$
If $\lambda_{(V,d)}(s,t)\geq\tau$, this implies
$$ \min\left\{\tau, \lambda_{(\cV,d_\cV)}([s],[t])\right\} = \min\left\{ \tau, \lambda_{(V,d)}(s,t)0\right\}. $$

Now assume that $\lambda_{(V,d)}(s,t) < \tau$. In this case, no set $S$ separating $s$ and $t$ with
$d(S,V\setminus S) = \lambda_{(V,d)}(s,t)$, can separate two elements $u$ and $v$ with $[u]=[v]$, 
because $\lambda_{(V,d)}(u,v) \geq \tau$. Hence each
partition $[u]$ is contained in either $S$ or $V\setminus S$. Therefore, $S$ induces a subset $\cS$ of $\cV$
with $\bigcup\cS = S$ and hence $\lambda_{(\cV,d_\cV)}([s],[t]) \leq  \lambda_{(V,d)}(s,t)$, implying
\begin{align*}
\min\left\{\tau, \lambda_{(\cV,d_\cV)}([s],[t])\right\} 
&= \lambda_{(\cV,d_\cV)}([s],[t]) 
\leq \lambda_{(V,d)}(s,t) \\
& = \min\left\{ \tau, \lambda_{(V,d)}(s,t)\right\} \leq \min\left\{\tau, 
\lambda_{(\cV,d_\cV)}([s],[t])\right\}
\end{align*}

The second part of the lemma is a direct consequence of the first part and the definition of $\lambda_{(V,d)}$.
\end{proof}

Now let $v_1,\dots, v_n$ be a lax-back order with threshold $\tau$ of $(V,d)$. Then define $\cV$
as the partition of $V$ consisting of the classes of the transitive and symmetric closure of the
relation $v_i \sim v_{i-1}$ iff $d(v_i,\{v_1,\dots, v_{i-1}\}) \geq\tau$.\footnote{Equivalently,
we may say, that the classes of $\cV$ are the maximal subsequences $v_i,\dots, v_k$ such that
$d(v_j,V_j)\geq\tau$ for $i<j\leq k$.} By lemma \ref{lem:triangle}, $\cV$ satisfies the condition of 
\ref{lem:ident} and hence
$$ \min\left\{\tau, \lambda_{(\cV,d_\cV)}\right\} = \min\left\{ \tau, \lambda_{(V,d)}\right\}.$$
This simple observation is the main tool for the algorithm presented in the next section.

\section{Minimizing Symmetric Set Functions}

\subsection{Construction of Lax-Back Orders}

\begin{algorithm}[t]\label{alg:lao}
\caption{Lax-Back-Order}
\BlankLine
\KwIn{A consistent symmetric set function $d$ on a finite set $V$, given by a lax oracle $F$ 
	and a threshold $\tau$}
\KwOut{A lax-back order $L=(v_1,\dots,v_n)$ with threshold $\tau$ for $(V,d)$\\\vspace*{-2mm}\hrulefill}
\KwData{An ordered list $L$ and a subset $U$ of $V$\\\vspace*{-2mm}\hrulefill}
\BlankLine
$L=(v_1)$\;
$U=V\setminus v_1$\;
\While{$|U|\geq 1$}{
	$x := 0$\;
	\ForAll(){$u'\in U$}{
		\uIf{$F(u,V\setminus U;\tau)\geq \tau$}{
			$L:=(L,u), U:=U\setminus u$\tcp*{Append $u$ to $L$.}
			$x:=\tau$\;
		}
		\ElseIf{$F(u,V\setminus U;\tau)>x$}{
			$x:=F(u,V\setminus U;\tau)$\;
			$u:=u'$\;
		}
	}
	\If{$x<\tau$}{
		$L := (L,u), U := U\setminus u$\tcp*{Append $u$ to $L$.}
	}
}
\end{algorithm}

Our algorithm for the calculation of a minimum bipartition of $(V,d)$, requires the construction
of a lax-back order as a subroutine. To achieve this, we assume, that we have access to a
{\em lax oracle} $F$ for $d$, ie. a program which returns $\min\left\{ \tau, d(S,T) \right\}$
for any input of two disjoint subsets of $V$ and a threshold $\tau$.

The correctness of the algorithm is intuitively clear. Nevertheless we will prove it in detail.
We show that, if an element $u$ is added to $L$ in lines 7 or 15, it satisfies the inequality
$$ \min \left\{ \tau, d(u,V\setminus U) \right\} \geq \min\left\{ \tau, d(u',V\setminus U) \right\} $$
for every $u'\in U$. Since $V\setminus U$ contains all elements already in $L$, this
implies that $L$ is an lax-back order with threshold $\tau$ for $(V,d)$ at the end
of the algorithm.

If $u$ is appended to $L$ in line 7, we have $d(u,V\setminus U)\geq \tau$ and
hence 
$$ \min \left\{ \tau, d(u,V\setminus U) \right\} = \tau \geq \min\left\{ \tau, d(u',V\setminus U) \right\} $$
for each $u'\in U$. 

If $u$ is appended to $L$ in line 15, no element was appended to $L$ in line 7 in this round.
Hence $d(u,V\setminus U)$ is obvious maximum among all $u'\in U$. This concludes the proof of
the correctnes of algorithm \ref{alg:lao}.

In each round of the while loop, exactly $|U|$ calls of the oracle are required.
Since in each round at least one element is removed from $U$, and since $U$ begins with
$|V|-1$ elements (line 2), at most $\frac{n(n-1)}{2}$ calls to the oracle are made with $n=|V|$.
Since it is possible, that more than one element is removed from $U$ in a round, this bound may be
very conservative, depending on the function $d$. As a result, a lax-back order can be calculated
in time $O(n^2 T_F)$, where $T_F$ is an upper bound for the execution time of the lax-oracle $F$.

But, depending on the lax oracle, the runtime required for the construction of a lax-back order may
be smaller. For example, as proven in \cite{stoer97mincut}, the construction of a {\em maximum adjacency order} on a weighted, undirected graph $G=(V,E)$, the analogue of a max-back order, requires time $	O(m+n\log n)$, where
$n=|V|$ and $m=|E|$. In \cite{brinkmeier05mincut} this was `reduced' to time $O(m+\delta n)$ for the
first, and $O(\delta n)$ for each subsequent lax-back order with a threshold $t\leq \delta$.
Here $\delta$ is the minimum degree of $G$.

As the above example indicates, it is possible in many situations to calculate $d(u, V_i)$
from $d(u,V_{i-1})$ much fater, than computing $d(u,V_i)$ directly. In these cases a variation
of algorithm \ref{alg:lao} in the fashion of the algorithm described in \cite{queyranne98minimizing}
is more apropriate.

In algorithm \ref{alg:lao2} we use a {\em priority queue $Q$ with threshold $\tau$}, ie.
$Q$ contains value-key-pairs $(v,k)\in V\times \mathbb{R}$ and provides three operations.
\begin{itemize}
\item The {\em insert} operation adds a pair $(v,k)$ to $Q$.
\item The {\em del\_max} operation removes a pair
$(v,k)$ from $Q$ such that $$k \geq \min\left\{\tau, \max\{k' \mid (v',k')\in Q\}\right\}.$$
\item The {\em update\_key} operation updates the key $k=d(u,V_i)$ to $k=d(u,V_{i+1})$ if
$v_i$ was extracted from $Q$.
\end{itemize}

The time required by algorithm \ref{alg:lao2} obviously depends on the type of thresholded
priority queue used and the time required by {\em update\_key}. But, since we are looking at
the most general case, we simply assume that each lax-back order can be calculated in time
$T_{LOB} \in O(n^2 T_F)$.

\begin{algorithm}[t]\label{alg:lao2}
\caption{Lax-Back-Order with Priority Queue}
\BlankLine
\KwIn{A consistent symmetric set function $d$ on a finite set $V$, given by a lax oracle $F$ 
	and a threshold $\tau$}
\KwOut{A lax-back order $L=(v_1,\dots,v_n)$ with threshold $\tau$ for $(V,d)$\\\vspace*{-2mm}\hrulefill}
\KwData{An ordered list $L$ and a priority queue $Q$ with threshold $\tau$.\\\vspace*{-2mm}\hrulefill}
\BlankLine
\lForAll{$v\in V$}{$Q.\text{insert}(v,-\infty)$}\;
\While{$|Q|\geq 0$}{
	$v := Q.\text{del\_max}()$\;
	$L:=(L,v)$\tcp*{Append $u$ to $L$.}
	\lForAll{$u\in Q$}{$\text{update\_key}(u)$}\;
}
\end{algorithm}

\subsection{Constructing a Minimum Bipartition}

Algorithm \ref{alg:mincut} constructs
a minimum bipartition, using algorithm \ref{alg:lao} as a subroutine. Before we begin with the analysis
of the algorithm, three remarks are appropriate:
\begin{itemize}
\item In the given form, algorithm \ref{alg:mincut} first builds a max-back order, since $\tau=\infty$.
Alternatively, the initial threshold may be set to 
$$ \min\left\{d(v,S\setminus v) \mid v\in V\right\},$$
resulting in a lax-back order, even in the first round. But in the worst case this requires
$n$ additional calls of the oracle.
\item Of course, we assume in lines 6, 8 and 11, that the $d\left([v_i],\left\{ [v_1],\dots
,[v_{i-1}]\right\}\right)$ were stored during the construction of the lax-back order.
\item We assume that lines 10-14 require $O(n)$ time.
\end{itemize}

\begin{algorithm}[t]\label{alg:mincut}
\caption{OptimalSet}
\KwIn{A consistent symmetric set function $d$ on a finite set $V$}
\KwOut{A subset $S$ of $V$ with $d(S,V\setminus S) = \lambda_ {(V,d)}$\\\vspace*{-2mm}\hrulefill}
\KwData{A partition $\cV$ of $V$ and a real number $\tau$\\\vspace*{-2mm}\hrulefill}
\BlankLine
$S := \emptyset$\;
$\cV := V$\;
$\tau := \infty$\;
\While{$|\cV|\geq 2$}{
	$([v_1],\dots, [v_k]) :=$ Lax-Back-Order$(\cV,d_\cV,\tau)$\;
	\If{$d([v_k],\cV\setminus [v_k]) < \tau$}{
		$S := [v_k]$\;
		$\tau := d([v_k],\cV\setminus [v_k])$\;
	}
	\For{$i=2, \dots, k$}{
		\If{$d\left([v_i],\left\{[v_1],\dots,[v_{i-1}]\right\}\right)\geq\tau$}{
			Join $[v_{i-1}]$ and $[v_i]$\;
		}
	}
}	
\end{algorithm}

Now we turn our attention to the correctness of algorithm \ref{alg:mincut}.
First, we prove that at the end of each execution of the body of the exterior loop (lines 5-14),
we have
\begin{equation*}
\tag{$\ast$} \lambda_{(V,d)} \leq\tau = d(S,V\setminus S) \quad\text{ and }\quad \forall u,v\in V\colon
[u]_\cV = [v]_\cV \Rightarrow \lambda_{(V,d)} \geq \tau,
\end{equation*}
where $ [u]_\cV$ denotes the equivalence class of $u\in V$ in $\cV$.

Observe that after the first execution of the body of the while loop,
we have $S=\{v_n\}$ and $\tau = d(v_n,V\setminus v_n)$. Hence ($\ast$) is satisfied.

Now assume that $S$, $\tau$ and $\cV$ are the `values' before the execution of the loop and
that they satisfy ($\ast$). If $d([v_k],\cV\setminus [v_k])\geq \tau$, the values of
$S$ and $\tau$ aren't changed and hence the first part of ($\ast$) is still valid at the end
of the loop-body.
If $d([v_k],\cV\setminus [v_k]) < \tau$ we obtain the new values
$\tau' = d([v_k],\cV\setminus [v_k]) < \tau$ and $S' = [v_k]$. Since $d_\cV([v_k],\cV\setminus [v_k])
= d(S',V\setminus S') = \tau'$, we obviously have
$ \lambda_{(V,d)} \leq \tau' = d(S',V\setminus S')$.

In line 10-14, sequences $[v_i]_{\cV},\dots, [v_j]_{\cV}$ of classes in $\cV$ are joined.
This results in a partition $\cV'$, such that each class $[u]_{\cV}$ of $\cV$ is contained
in a class of $\cV'$. 

Now assume that $[u]_{\cV'} = [v]_{\cV'}$ for two elements $s,t$ of $V$.
If $[u]_\cV=[v]_\cV$, we have $\lambda_{(V,d)}(u,v) \geq \tau$, since ($\ast$) was satisfied before
the sets were joined.

If $[u]_\cV\neq[v]_\cV$, we assume that $S$ separates $u$ and $v$ with $u\in S$ and
$\lambda_{(V,d)}(u,v) = d(S,V\setminus S)$. If $S$ also separates two elements $s$ and $t$
with $[s]_\cV = [t]_\cV$, we have $d(S,V\setminus S) \geq\tau$, by induction.

If on the other hand, each class of $\cV$ lies either in $S$ or in $V\setminus S$, w.l.o.g.
we have a sequence
$[u]_\cV=[v_i]_\cV, \dots, [v_j]_\cV=[v]_\cV$ of classes in $\cV$, such that
$$d([v_l]_{\cV},\left\{[v_1]_{\cV},\dots,[v_{l-1}]_{\cV}\right\})\geq\tau \text{ for } i< l \leq j.$$
Therefore, lemma \ref{lem:sub} implies
$$ \lambda_{(\cV,d_\cV)}([v_l]_\cV,[v_{l-1}]_\cV) \geq\tau \text{ for } i<l\leq j. $$
By lemma \ref{lem:ident} this implies 
$$\lambda_{(V,d)}(v_l,v_{l-1}) \geq \tau \text{ for } i< l \leq j$$
and hence, by lemma \ref{lem:triangle}, $\lambda_{(V,d)}(u,v) \geq\tau$, proving the
validity of ($\ast$) at the end of the loop.

Since $\tau \leq d_{\cV,d_\cV}([v_k]_\cV, \cV\setminus [v_k]_\cV)$ after line 9, at least
the sets $[v_{k-1}]_\cV$ and $[v_k]_\cV$ are joined in lines 10-14. Hence, the number of
partitions in $\cV$ decreases each round and after at most $|V|-1$ rounds, we have $|\cV|=1$,
and the algorithm terminates.

In this situation ($\ast$) implies $\lambda_{(V,d)} \leq \tau = d(S,V\setminus S)$
and since $[u]=[v]$ for all $u,v\in V$, we have $\lambda_{(V,d)}(u,v) \geq \tau$,
leading to
$$ \lambda_{(V,d)} = \tau = d(S,V\setminus S). $$

\begin{theorem}
Given a lax oracle $F$ for a monotone and consistent symmetric set function $d$ on $V$
\begin{enumerate}
\item The
algorithm \ref{alg:mincut} returns a set $S$ of $V$ with $d(S,V\setminus S)=\lambda_{(V,d)}$ in time
$O(n^3 T_F )$, where $T_F$ is an upper bound for the time required by the oracle $F$.
\item If $T_{LBO}$ is an upper bound for the construction of a lax-back order with
threshold $\tau\leq \min\left\{ d(v,V\setminus v)\mid v\in V\right\}$ for $(V,d)$, then
a set $S$ with $d(S,V\setminus S)=\lambda_{(V,d))}$ can be constructed in time
$O(n (T_F + T_{LBO}) )$.
\end{enumerate}
\end{theorem}


%
%
%
%


\section{Conclusion}

We observed that for each monotone and consistent symmetric set function $d$ on a finite set $V$,
the function $\hat d(S,T) = \min\left\{\tau, d(S,T)\right\}$ is monotone and consistent, too.
This fact was used to weaken the conditions on the oracle required for the construction of a
set $\emptyset \subset S \subset V$, minimizing $d(S,V\setminus S)$. Instead of using a
(strict) oracle $F$, we only required a {\em lax oracle} providing $\min\{\tau, d(S,T)\}$ for
a given threshold $\tau$. This allowed several improvements:
\begin{itemize}
\item Depending on $d$, a lax oracle may require less time than a strict oracle (an example can be found 
in \cite{brinkmeier05mincut}).
\item The usage of {\em lax-back orders} instead of {\em max-back orders} and the fact
that $d$ is monotone, allows a possibly faster construction of the order, since the number of oracle calls
per element is reduced.
\item Since more than one identification is possible per round, the total number of rounds may be reduced.
\end{itemize}

Unfortunately, all three improvements heavily depend on the function $d$. Hence, only a detailed
analysis of special cases, may lead to a guaranteed improvement. In full generality, we can only
guarantee, that the runtime of the algorithm presented in this paper is at least as fast as the
one in \cite{rizzi00symmetric}.

\bibliography{../bib/bibliography}
\bibliographystyle{alpha}

\end{document}